**Unipolar Resistance Switching in Amorphous High-*k* dielectrics Based on Correlated Barrier Hopping Theory****

By *Kui Li*, *Yidong Xia,** *Bo Xu*, *Hongxuan Guo*, *Xu Gao*, *Kang Guo*, *Jiang Yin*, *and Zhiguo Liu*

[*]  Dr. Y. D. Xia, to whom correspondence should be addressed
Department of Materials Science and Engineering, and National Laboratory of Solid State Microstructures,
Nanjing University, Hankou Road 22, Nanjing 210093, People's Republic of China
E-mail: xiayd@nju.edu.cn
      K. Li, B. Xu, Dr. H. X. Guo, K. Guo, Prof. Z. G. Liu
Department of Materials Science and Engineering, and National Laboratory of Solid State Microstructures,
Nanjing University, Hankou Road 22, Nanjing 210093, People's Republic of China
         X. Gao, Prof. J. Yin
Department of Physics, and National Laboratory of Solid State Microstructures,
Nanjing University, Hankou Road 22, Nanjing 210093, People's Republic of China

[**]This work was supported by a grant from the National Natural Science Foundation of China under a grant No. 10804048, the State Key Program for Basic Research of China (2010CB630704), Research Fund for the Doctoral Program of Higher Education of China (200802841003), 863 Project of China with a Grant No. 2008AA031401, and National Key Project (2009ZX02023-5-4).

**Abstract**: We have proposed a kind of nonvolatile resistive switching memory based on amorphous $LaLuO_3$, which has already been established as a promising candidate of high-*k* gate dielectric employed in transistors. Well-developed unipolar switching behaviors in amorphous $LaLuO_3$ make it suited for not only logic but memory applications using the conventional semiconductor or the emerging nano/CMOS architectures. The conduction transition between high- and low- resistance states is attributed to the change in the separation between oxygen vacancy sites in the light of the correlated barrier hopping theory. The mean migration distances of vacancies responsible for the resistive switching are demonstrated in nanoscale, which could account for the ultrafast programming speed of 6 ns. The origin of the distributions in switching parameters in oxides can be well understood according to the switching principle. Furthermore, an approach has also been developed to make the operation voltages predictable for the practical applications of resistive memories.

# 1. Introduction

Semiconductor nonvolatile memory cell generally consists of memory element and transistor element.[1] Both the silicon-based complementary metal oxide semiconductor (CMOS) and the conventional memory technologies have been faced with the scaling issues as the semiconductor devices are rapidly approaching the miniaturization limits.[2] The development of high-*k* (high dielectric constant) materials to replace silicon oxides as alternate gate dielectrics is the key solution to attain continuous CMOS scaling.[2-4] Meanwhile, novel memory concepts totally different from that based on charge-storage have emerged to construct the next-generation nonvolatile memories. Resistive switching memory (RRAM) as one of the promising candidates has recently triggered scientific and technological attentions.[1,2,5-17] The basic principle of RRAM is that the resistance of a metal-insulator-metal (MIM) device can be electrically modulated in a simple way.

A wide range of binary and multinary oxides have been demonstrated to show resistance switching.[6-20] Switching is termed as unipolar when the resistance switching procedure is independent of the polarity of the applied voltages,[6,7] which is of particular concern as it can offer much larger resistance change in simplified circuit architecture.[8,12] Most binary transition-metal oxides, in particular NiO and $TiO_2$,[8,9,12,16,20] exhibit unipolar resistance switching. The critical issue as to such unipolar behavior in these oxides is the understanding of switching mechanisms. The popular model proposed by far is the formation and rupture of conductive filament.[6,7,9,12,16,20] As the materials employed in MIM cell are usually in polycrystalline structure, grain boundaries can enable the development of filament. Nanofilament composed of metallic nickel at the grain boundary has been identified in NiO via transmission electron microscopy method.[12]

Another matter that should be noted for the polycrystalline case, however, is that the cell size will become comparable to the grain diameter as the industry moves toward 22-nm technology projected for 2016. This may cause detrimental device-to-device variations in switching characteristics due to diverse grain boundary nature in different memory cells. Amorphous materials free from grain boundaries are capable of offering homogeneous structure to avoid such issue. It is well reasoned that amorphous high-$k$ gate dielectrics, which have already been demonstrated to be compatible with semiconductor transistor technologies, can be good choices for RRAM applications as long as such materials can show well-developed resistance switching behaviors. Lanthanum based high-$k$ oxides are considered as the candidates beyond the hafnium-based technology,[3,4,21] among which LaLuO$_3$ is the promising one as it has high dielectric constant and large band gap, and can remain amorphous up to 1000 $^o$C.[21] We herein reveal unipolar resistance switching in Pt/amorphous LaLuO$_3$/Pt capacitors, and a model responsible for such switching in amorphous high-$k$ oxides has been proposed in the light of the correlated barrier hopping (CBH) theory. Relations between low resistances ($R_L$) and switching voltages have also been derived based on the statistic analyses aiming at the predictability of the programming functions.

**2. Results and Discussion**

A representative current-voltage (*I-V*) characteristic of Pt/amorphous LaLuO$_3$/Pt capacitor is illustrated as a semilogarithmic plot in Figure 1a. The capacitor stays initially in an insulating state until the applied voltage reaches a forming voltage, where a sudden increase in current takes place (inset of Figure 1a). A compliance current (CC) value is set to limit the current to protect the device from dielectric breakdown. The device goes into a high conductive state, as the current is rather high even at small voltages (curve B). Once the voltage is greater than a RESET value ($V_R$), the current drops abruptly, indicating the device is tuned back to a high resistance state (HRS) from the high conductive one. This HRS is stable in the voltage range below a SET voltage ($V_S$), beyond which the current increases suddenly (curve A) to the CC value and the device switches to the low resistance state (LRS) once again. Such bistable switching displays the typical unipolar manner similar to those observed in NiO and $TiO_2$.[8,9,12,16,20]

High programming speed is one of the attractive merits for RRAM to replace the conventional charge-storage technology.[9,11,12] The programming speed per cell in Flash memory nowadays is on the order of tens of microseconds,[9] whereas RRAM cells based on $LaLuO_3$ can be SET as fast as 6 ns. The speed measurements are performed via pulse switching mode. We first test the device at HRS with smaller DC bias (from 0 to 0.1 V) before the application of SET pulse, where the low current values confirm the device at HRS (curve I in Figure 1c). A 6.5 V pulse with 6 ns duration is then applied to SET the device (as shown in Figure 1b), after which another small signal DC test is used to verify the LRS (curve II in Figure 1c). The RESET speed is also obtained via the above steps and typical switching time is verified to be < 60 ns. Moreover, the switching time for SET process exhibits dependence on the SET voltages. The smaller $V_S$ is, the longer switching time will be, as indicated in Figure 1d. Note that 6 ns is the limitation for our pulse system, it can be expected that the memory cell can be turned on in less than 6 ns by proper SET pulse. These desirable memory performances make the $LaLuO_3$ well-suited for both memory and logic applications compatible with the conventional or the emerging nano/CMOS architectures.[10]

The retention and endurance characteristics at room temperature are executed to evaluate the nonvolatile performances, as illustrated in Figures 2a and 2b, respectively. The ratios between HRS and LRS can be held greater than $10^7$ without any indication of degradation within the measurement duration under a small reading voltage of 0.1 V. The bistable resistances as a function of switching cycles are summarized in Figure 2b in 2200 successive cycles, where memory window can be clearly distinguished. However, fluctuations in HR and LR also occur, as well as the distributions of $V_S$ and $V_R$, which are the major obstacles to practical applications for oxide RRAMs.[8,22] Figure 2c shows the statistic results of $V_S$ and $V_R$. $V_R$ is rather stable as indicated by the narrow voltage dispersion between 0.3 and 1.2 V, whereas $V_S$ exhibits wide distribution from 1.6 to 13 V. In spite of the distributions in the switching voltages, no overlap of $V_R$ and $V_S$ is detected, which offers a clear window for the programming voltages to avoid the irresolvable errors in RRAM functions. Efforts have been put forth to make the switching operations controllable or predicable.[8,22] However, by now it seems that such fluctuation is rather an intrinsic matter, which might be correlated to certain switching mechanism. As a result, it is yet of importance to make clear the origin of the unipolar switching in oxides.

Several conduction mechanisms control the *I-V* characteristics in oxide films.[19] It is worthwhile to comprehend the switching with regard to the conduction transition between HRS and LRS. In most conduction mechanisms for MIM systems the current *I* is exponentially dependent on the electric field *E*. A powerful method for exploring *I-V* relationships to determine the dominant contribution mechanisms has been presented by Niklasson and Brantervik,[23] which is based on the analyses of the derivative of the logarithmic conductivity with respect to inverse applied electrical field. According to this method, a quantity $\Delta$ is defined to figure out the conduction mechanism as

$$\Delta \equiv \frac{d(\ln \sigma)}{d(1/E)} \approx nAE^{1-n}, \qquad (1)$$

where $\sigma$ is the conductivity, *A* is a possibly temperature-dependent quantity, and different values of parameter *n* refer to various conduction mechanisms. Figures 3a and 3b present the plots of $\log \Delta$ against $\log E$ for the capacitors set at HRS and LRS respectively, by which the slopes of $1-n$ are obtained to be 2 (for HRS) and 3 (for LRS) respectively. Accordingly, $n = -1$ for HRS corresponds to the so-called Poole's law. Poole's law is the mechanism that the conduction is determined by the transport of electrons from donor levels and the overlap of potential wells takes place in the case of a high density of ionized donors.[23] On the other hand, the conduction in LRS is dominated by percolation with $n = -2$.[23]

Metallic-like conduction behavior in LRS is furthermore verified via temperature dependence of resistances, as shown in Figure 4a. Resistance increases slightly with the increasing temperature from 25 to 300 K. Evidence has confirmed the metallic precipitates forming the conductive paths in LRS, such as metallic nickel in NiO RRAM.[12,16,20] Nevertheless, things are different in LaLuO$_3$, as no metallic bonding phases can be detected in LRS. Depth profile of x-ray photoelectron spectroscopy (XPS) of La 3d and Lu 4d are shown in Figure 4b, respectively. Both La and Lu in LRS are identified in the +3 valence states without any metallic or reduced bonding signal, according to the detected La 3d$_{5/2}$ and Lu 4d$_{5/2}$ peaks located at 833.8 and 194.9 eV, respectively.[24]

Now that chemical metallic phases are absent in LaLuO$_3$ capacitors, it is necessary to clarify the origin of such metallic conduction behavior. We herein convey the conduction transition in terms of the CBH theory. Based on the CBH theory first introduced by Pike,[25-27] the charge carrier is assumed to hop between site pairs over the potential barrier separating them. For simplification, if a single electron hopping between positive defect centers is considered then the barrier height $W$ is correlated with the separation $r$ according to

$$W = W_m - \frac{e^2}{\pi \varepsilon \varepsilon_0 r}, \qquad (2)$$

where $\varepsilon$ is the dielectric constant of matrix material, and $W_m$ is the energy difference between the potential well of localized states and the band (or extended) state. This expression shows the dependence of barrier height $W$ on the separation $r$. The closer distance between localized states (i.e. smaller $r$) results in the reduced barrier height and hence facilitates the electron transport in the materials. It is hence valid to expect that the smaller (larger) separation $r$ is, the lower (higher) resistance will be.

It is well known that oxygen vacancies play critical role in the conduction in oxides.[17-19,28,29] The conduction transition between HRS and LRS is consequently attributed to the electron transport controlled by oxygen vacancies. Considering the migration of oxygen vacancies driven by the applied electric field, the observed unipolar resistive switching can then be pictured in the light of the CBH theory. For convenience, the subscripts H and L in the following sections stand for the HRS and LRS. At HRS the localized oxygen vacancies are considered separated by a larger distance $r_H$ leading to the higher barrier $W_H$ to limit the hopping between them, which makes the conduction follow the Poole's law. When a higher electric field is applied onto the MIM structure, more oxygen vacancies are collected and hence form local regions with higher density of defect states.[30,31] Overlap of potential wells can be expected in this process. Higher defect density means smaller distance between the vacancy sites, which reduces the barrier height low enough to make the electrons transport between the localized states more easily and accordingly enhances the conduction remarkably, corresponding to the SET process. LRS is thus established, and the resistance mainly comes from the scattering by the vibration of vacancies, where conduction exhibits metallic-like behavior as indicated in Figure 4a.

The change in the site separation responsible for the resistance transition can be estimated according to Eq. 2. $W_m$ is assumed constant in the two resistance states so that we can simply discuss the relationship between the two variables, $W$ and $r$. The energy level of oxygen vacancies in orthorhombic-perovskite $LaLuO_3$ is first computed via density functional theory (DFT) within local-density approximation method. The total density of states (DOS) and the partial DOS of O are plotted in Figure 5a. According to the calculation, the energy level of oxygen vacancy lies in the oxide band gap and below the conduction band edge with about 1.05 eV, which corresponds to the $W_m$ in Eq. 2. The second step is the determination of the barrier height $W$. In the case of HRS, the activation energy $E_a$ for conductivity is taken as the $W$ for electrons hopping between the vacancy sites. Figure 5b shows the temperature dependence of conductivity, which is derived under a reading bias of 0.1 V. According to the Arrhenius relationship of[11,29]

$$\sigma = \sigma_0 \exp\left(-\frac{E_a}{kT}\right), \tag{3}$$

where $\sigma_0$ is the preexponential factor, $k$ is Boltzmann's constant, and $T$ is the absolute temperature, the activation energy $E_a$ of 0.91 eV can be drawn from the slope of the linearly fitting line. $r_H$ is then calculated to be 1.28 nm with $\varepsilon$ of 32 (taken from Ref. [21]). Similar treatment is executed for LRS except the difference in $W$. As the metallic-like conduction observed in LRS, we here take the approximate condition of $W_L = 0$ for the electron transport between sites, resulting in the minimum $r_L$ of 0.18 nm. This indicates that the change of site separation responsible for the SET process is localized in nanoscale (approximately 1 nm). Here it is worth pointing out that the oxygen vacancies are not necessarily aligned in filamentary configurations throughout the capacitor to structure the LRS. These defects will be arranged in network feature as long as the majority of vacancy sites are close sufficiently to form the LRS, as depicted in the schematic diagram in Figure 5d. The formation dynamics of conducting network in the SET process has not been understood well. One of the possible mechanism is the soft electric breakdown.[32] However, due to the lack of data, we can only analyse the formation of such network qualitatively. The clustering of these defects should result from two competing processes in energy. One is the decrease in Gibb's free energy during the clustering under the electric field. The other is the increase in energy from the Coulomb repulsion when the vacancies get much closer. Since the migration of vacancies necessary for building such conductive network is localized, the smaller migration distance makes the high SET speed (< 6 ns) feasible. Once a $V_R$ is applied to the LRS capacitor, the current is so high that the resultant Joule heat ruptures the network and drives the vacancies more apart to longer separation $r_H$. The capacitor is in consequence switched back to the HRS, corresponding to the RESET process.

According to the above discussions, the origin of distributions of switching parameters can be understood well. As the transition between resistances is owing to the change in the separation between defect sites in nanoscale, the variation of resistance values stems from the fluctuation of vacancy-migration distances. In RESET process, Joule heat disperses the assembled vacancies leading to the rupture of the correlated pairs. But not all the site pairs will be broken up in this process. Once the fraction of ruptured pairs exceeds the threshold, similar to the case in random circuit breaker model (Refs. [8] and [33]), the current drops remarkably and thereby no sufficient Joule heat can be produced to drive the vacancies more apart and further rupture the rest site pairs. The degree of the remained correlated sites determines the value of $R_H$. Lower $R_H$ is related to smaller $r_H$ or more correlated pairs remained. Variations of $R_H$ can hence be expected, as already shown in Figure 2b. In SET process, the scale of $R_H$ can affect the switching voltages. Figure 6a presents the statistic results of $V_S$ against $R_H$ within 216 successive switching cycles. The trend line marked in the figure indicates that the higher $R_H$ corresponds to the larger SET voltage. Higher $R_H$ means the larger $r_H$ or the fewer correlated pairs remained, which makes the SET process more difficult. That is why larger SET voltage is necessary.

As the RESET process is Joule heat controlled, certain correlation between the RESET power and $R_L$ can be expected.[22,34] Figures 6b and 6c show the statistic results of $R_{L0}$ dependent $I_R$ and $V_R$, respectively. Here $R_{L0}$ is the resistance value tested under a reading bias of 0.1 V. The data fitting reveals that both $I_R$ and $V_R$ are proportional to $R_{L0}^{-\alpha}$ and two scaling regimes exist in both plots:

$$\begin{cases} I_R \propto R_{L0}^{-2.69} (R_{L0} < R_T) \\ I_R \propto R_{L0}^{-1.30} (R_{L0} > R_T) \end{cases}, \quad (4)$$

and
$$\begin{cases} V_R \propto R_{L0}^{-1.55} (R_{L0} < R_T) \\ V_R \propto R_{L0}^{-0.36} (R_{L0} > R_T) \end{cases}, \quad (5)$$

where $R_T$ of 14 Ω is the critical value for the two scaling regimes. These two relations provide evidence of percolation conduction with multiply connected network in LRS.[34,35]

Furthermore, these scaling relations are valid for other amorphous high-$k$ dielectrics that also exhibit unipolar switching, such as $Hf_{1-x}Si_xO_4$ systems (data not shown here).

Since the presence of distributions in switching parameters is inevitable, the predictability of operation voltages is rather of importance. Lately, some methods have been carried out to predict the RESET voltages,[8,22] but little progress has been made on SET issues. In view of this, an approach is developed to predict the operation voltages. Equation 5 can be employed to make the RESET voltages predicable. As $R_{L0}$ can be measured via a reading bias, the value of $V_R$ is thus obtained. As indicated in Figure 2c, $V_S$ exhibits more wide distribution than $V_R$ does. However, no exact relationship can be drawn directly between $V_S$ and $R_H$ (as shown in Figure 6a), making the predictability of SET voltages more challenging. Instead of dealing with the $V_S$ against $R_H$, we here analyze the correlation between $V_S$ and $V_R$. The product of $V_R$ and $V_S$ is plotted according to $V_S$ in Figure 6d, and the fitting function can be expressed as

$$V_R V_S = C V_S^{1.5}, \qquad (6)$$

where $C$ is the fitting constant. $V_S$ will be consequently estimated as long as the $V_R$ is determined from Eq. 5. Resulting values of the simulated $V_R$ and $V_S$ based on these empirically observed relations are summarized in Figure 6e. The comparison of the predicted and experimental data demonstrates the feasibility of our proposed approach, although further reduction of fitting errors should be improved necessarily. This scenario could be universal and applicable to other RRAMs based on amorphous oxides. A detailed analysis of these relationships to further understand the mechanism will be the subject of a future study.

## 3. Conclusions

In conclusion, the well-developed unipolar resistive switching revealed in amorphous $LaLuO_3$ capacitors make it suited for both logic and memory elements in nonvolatile memories, which facilitates the integration and compatibility with the progressive semiconductor technologies. Considering that cell size dependence of switching current, power and time has been reported in several studies,[9,12] improvement of the memory performance with further scaling of LaLuO$_3$ cell size can be evaluated accordingly. The transition of resistances is attributed to the change in the separation between oxygen vacancy sites in nanoscale in terms of the CBH theory, which could account for the ultrafast programming speed of 6 ns and is of importance for understanding the origin of the distributions in switching parameters. In addition, based on the statistic analyses, an approach has also been proposed to predict the operation voltages to solve the reliable issues, which is essential for the practical applications of oxide RRAMs.

**4. Experimental**

*Experimental details and electrical measurements of the LaLuO$_3$ memory devices*: The LaLuO$_3$ films studied in this work have been derived by pulsed laser deposition method using a KrF excimer laser (COMPex, Lambda Physik, 248 nm in wavelength, 30 ns in pulse width). LaLuO$_3$ ceramic target was fabricated via standard milling, cold pressing and sintering processes with the mixed powder of Lu$_2$O$_3$ and La$_2$O$_3$ (mol. ratio of 1:1). LaLuO$_3$ films with the thickness of ~ 150 nm were grown on the commercially available Pt/TiO$_2$/SiO$_2$/Si substrates at the fixed temperature of 400 °C with chamber vacuum of $2\times10^{-4}$ Pa. The MIM capacitor structures were fulfilled after the deposition of Pt top electrodes with the diameter of 0.2 mm, which are deposited through a shadow mask onto the films at room temperature by dc magnetron sputtering. The samples were then annealed at 500 °C in N$_2$ for 5 mins. The amorphous nature of the as-deposited films as well as the annealed samples was confirmed by x-ray diffraction using a D/Max-rA diffractometer (Rigaku, Japan) with Cu K$\alpha$ radiation (data not shown here). The chemical states of LaLuO$_3$ capacitors set at LRS were analyzed by XPS using Thermo ESCALAB 250 system, equipped with a monochromatic Al $K\alpha$ (1486.6 eV) source and pass energy of 20 eV. Ar$^+$ sputtering was used to obtain the depth profile. Narrow-scan spectra of all interested regions (film bulk, as well as the top and bottom electrode/film interfaces) have been recorded to identify chemical binding states of the corresponding elements. *I-V* behaviors were performed by using a Keithley 2400 source-measure unit. The programming speed was executed via pulse switching mode, employing a function generator (Agilent 81104A) and a digital storage oscilloscope (LeCroy WaveRunner 62Xi). The activation energy $E_a$ for conductivity was derived from the fitting slope of temperature dependence of high resistances from 413 to 533 K within a temperature controlled probe station. The low resistances as a function of temperature from 25 to 300 K were performed by physical properties measurement system (PPMS) to verify the metallic conduction behavior in LRS.

*DFT calculations*: The calculations are implemented in VASP (Vienna *ab initio* simulation package) code using projector-augmented-wave pseudopotentials with the cutoff energy of 500 eV for plane wave expansion of the electronic wave function. The electron structure and total energy calculations are performed with a 40-atoms supercell containing a single oxygen vacancy. The Brillouin zone integrations are performed on a well converged grid of $8 \times 6 \times 4$ *k*-points. The total energy is converged to $1 \times 10^{-5}$ eV/atom. Good convergence is obtained with these computing parameters.

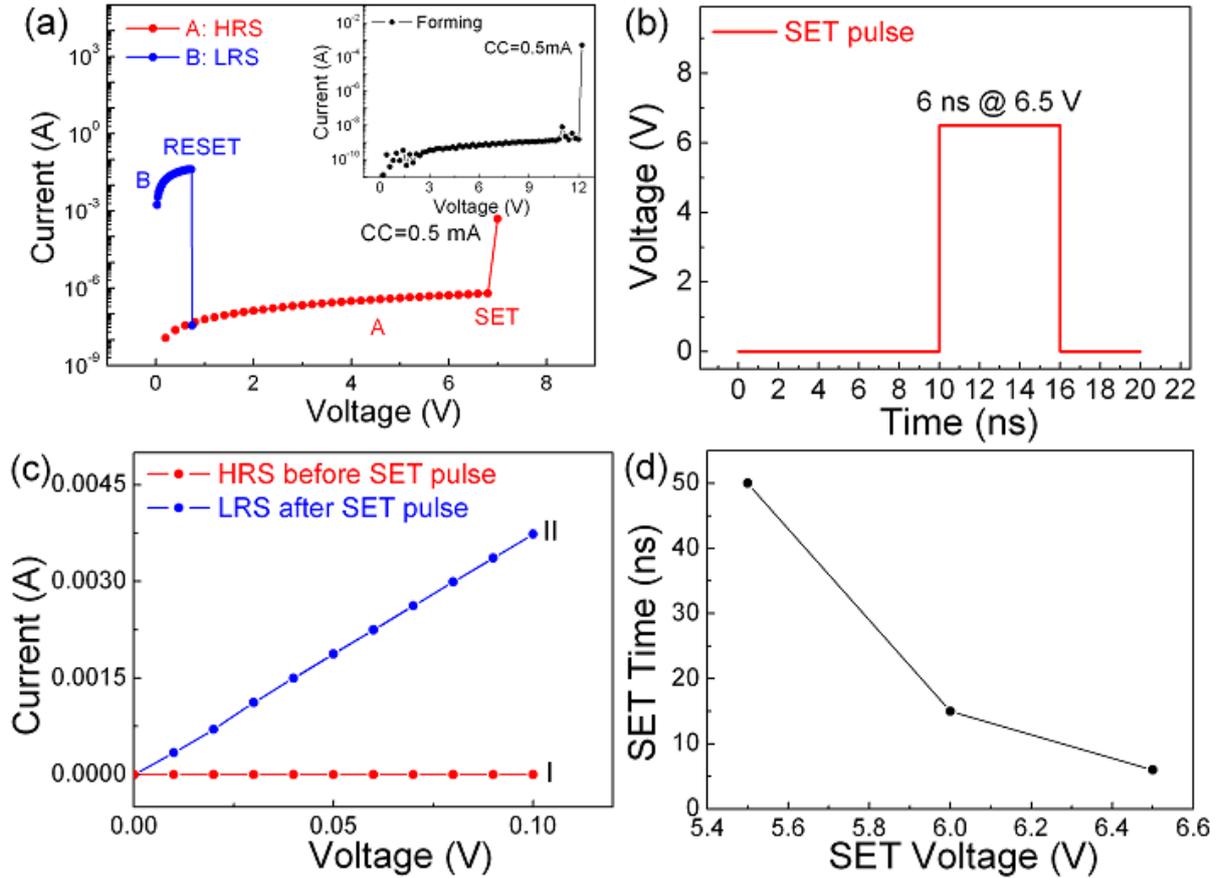

**Figure 1.** Unipolar resistance switching in Pt/amorphous LaLuO$_3$/Pt capacitors. (a) Semilog plot of a typical *I-V* characteristic, indicating the unipoalr behavior. Curves A and B correspond to the HRS and LRS, respectively. The inset shows the forming process. (b) Measurements of switching speed via pulse switching mode. A 6.5 V SET pulse with 6 ns duration. (c) Reading processes with small bias signal from 0 to 0.1 V to confirm the resistance change before (curve I) and after (curve II) the application of SET pulse. (d) The dependence of switching time on SET voltage.

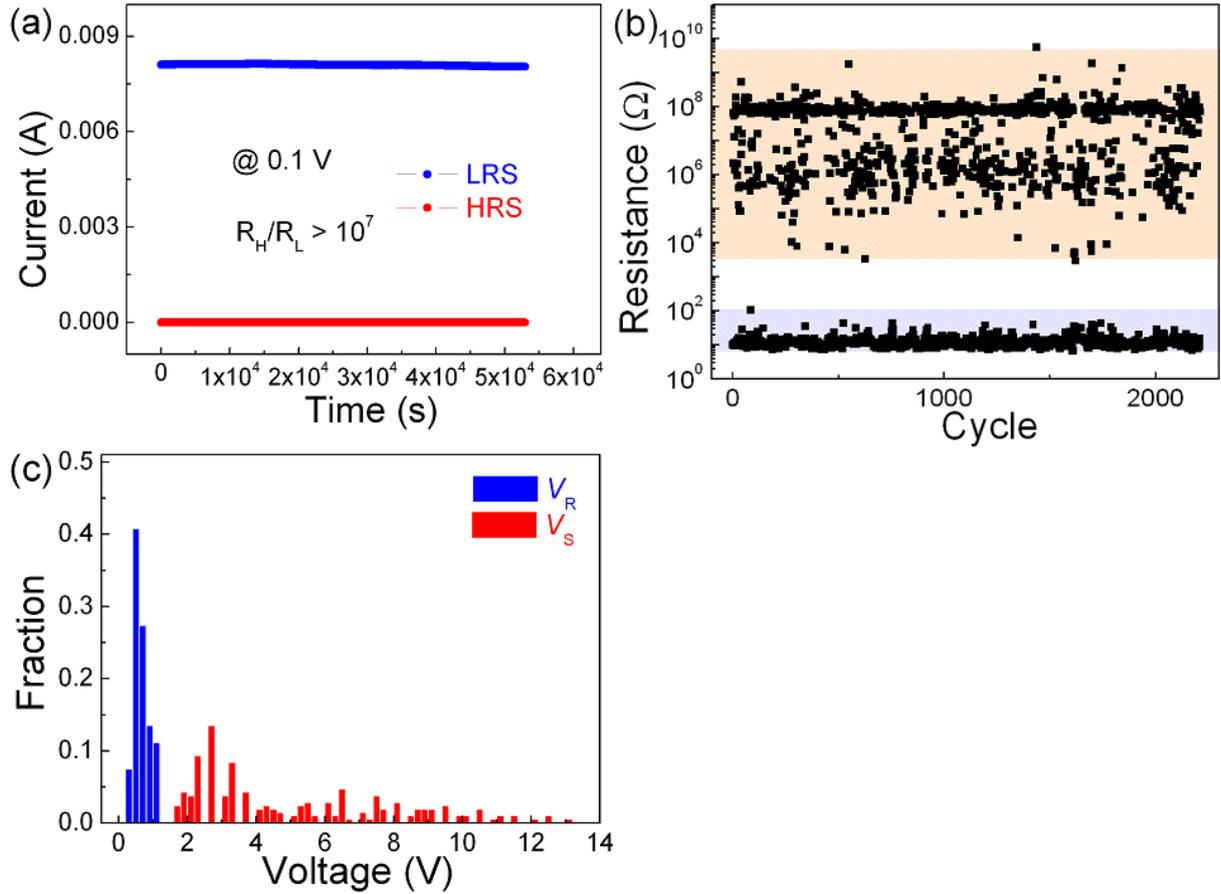

**Figure 2.** Reliability of memory performances at room temperature. (a) Retention characteristic measured under a reading bias of 0.1 V. (b) Bistable resistances as a function of switching cycles in 2200 successive cycles, indicating the fluctuations in HR and LR. (c) Distributions of $V_S$ and $V_R$ in 216 successive cycles from (b).

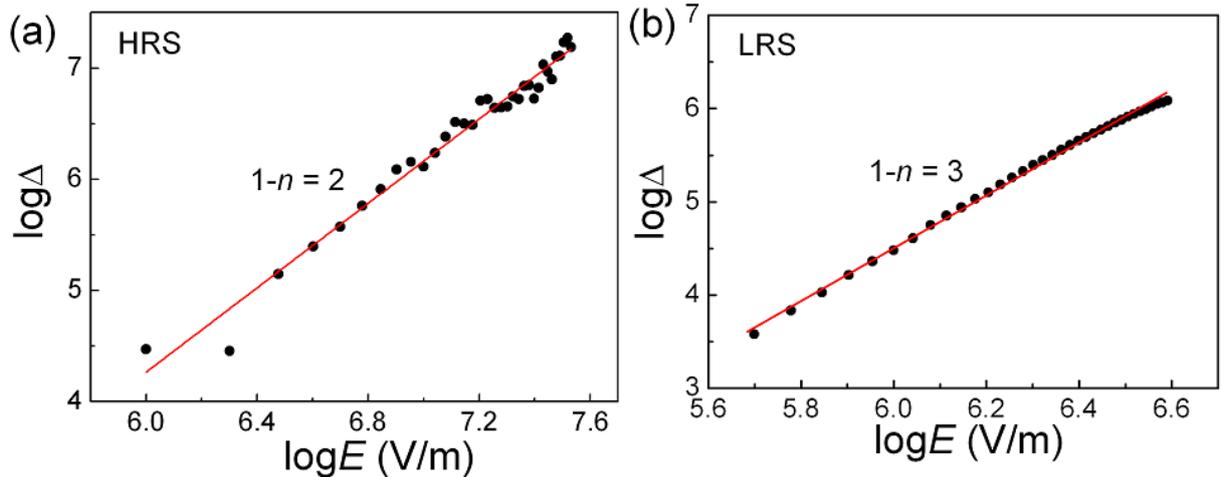

**Figure 3.** Determination of conduction mechanisms for (a) HRS and (b) LRS according to the slopes of linear relationships between $\log \Delta$ and $\log E$, respectively. $\Delta$ is calculated using Eq. 1.

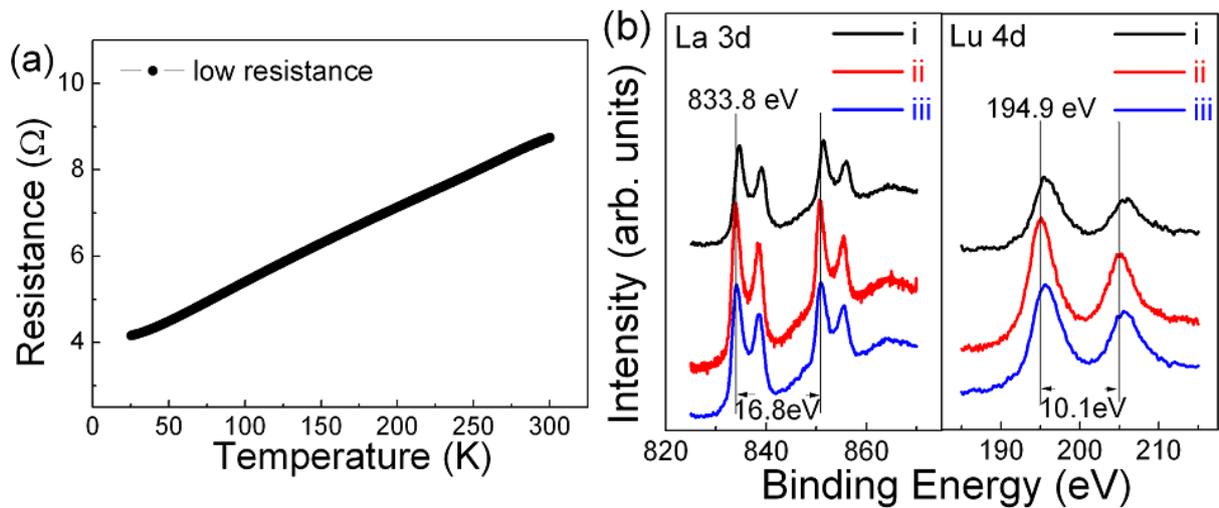

**Figure 4.** Metallic-like conduction in LRS. (a) Temperature dependence of resistances in LRS from 25 to 300 K. (b) Depth profile of XPS for La 3d and Lu 4d in LRS. Curves i, ii, and iii correspond to the detected regions of interface at top electrode/film, film bulk, and interface at film/bottom electrode, respectively. No metallic or reduced bonding states can be detected.

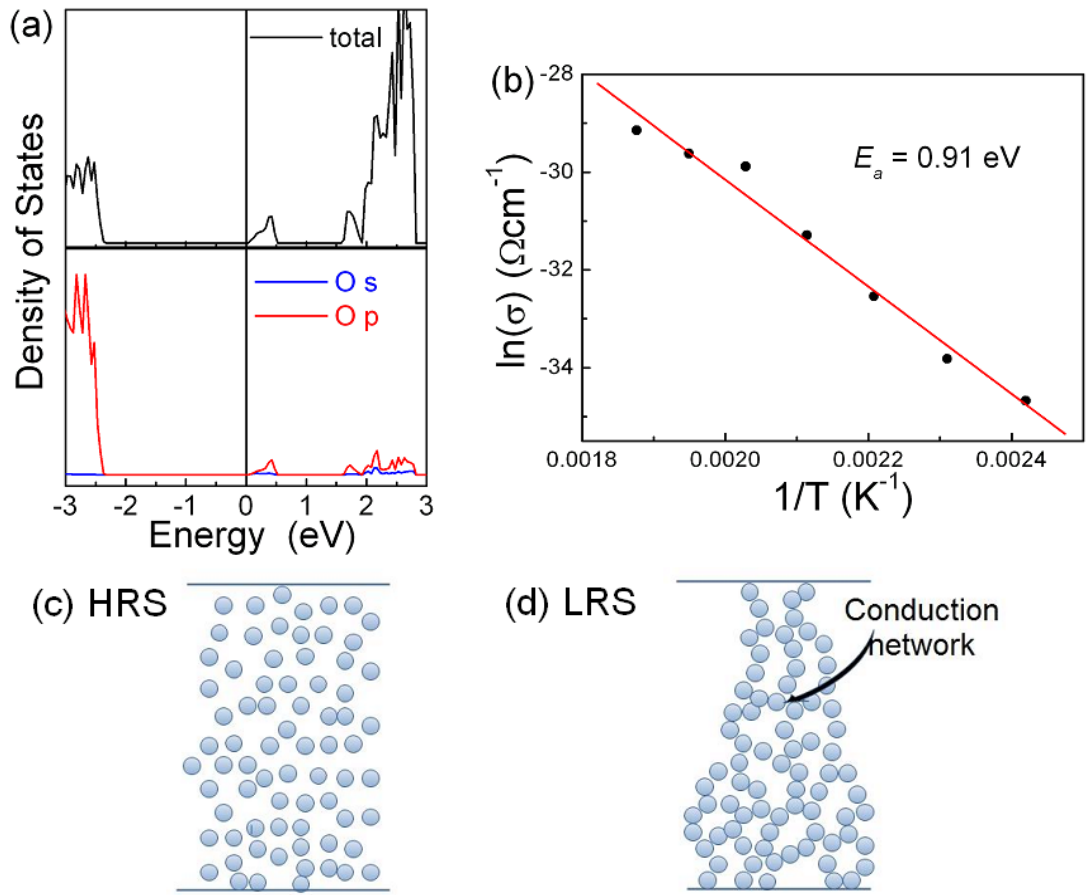

**Figure 5.** Proposed mechanism responsible for the unipolar switching in amorphous LaLuO$_3$ capacitors. (a) DFT calculation of the energy level of oxygen vacancy. (b) Arrhenius-type temperature dependence of conductivity in HRS, from which the activation energy $E_a$ of 0.91 eV is derived. Schematic diagrams for the configurations of oxygen vacancies in (c) HRS and (d) LRS, respectively.

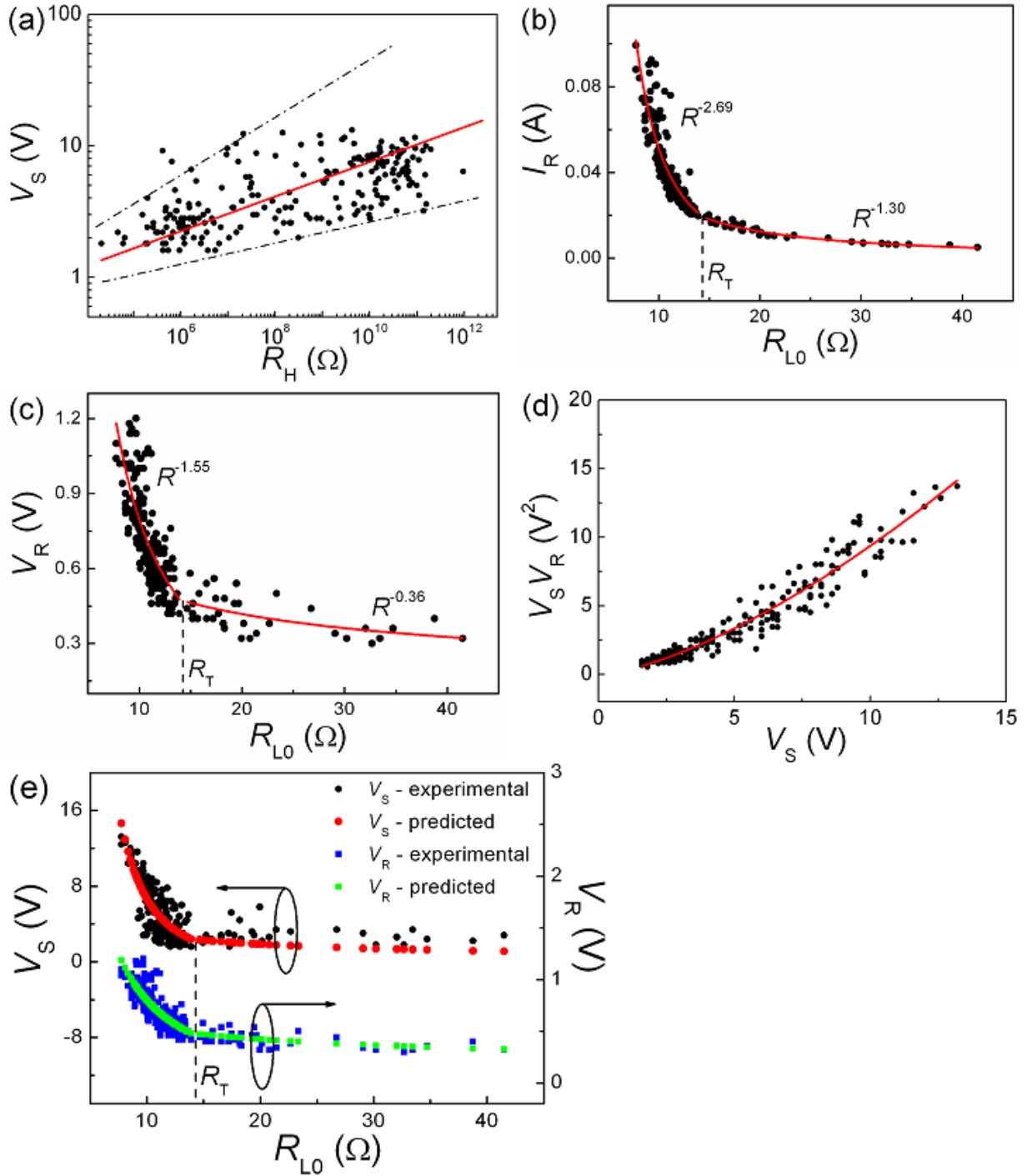

**Figure 6.** Predictability of switching voltages. (a) $V_S$ versus $R_H$ within 216 successive switching cycles. The dashed lines indicate the upper and lower bounds of the data. The red marked trend line indicates that the higher $R_H$ corresponds to the larger SET voltage. (b) The dependence of $I_R$ on $R_{L0}$, showing the proportional relationship between $I_R$ and $R_{L0}^{-\alpha}$. The data fitting reveals the existence of two scaling regimes. (c) The dependence of $V_R$ on $R_{L0}$, from which the $V_R$ can be predicted with the measured $R_{L0}$. (d) Plot of the product of $V_R$ and $V_S$ against $V_S$, from which the $V_S$ can be determined. All the fitting curves are red marked in (b), (c) and (d). (e) The comparison of the predicted and experimental data. The predicted $V_R$ and $V_S$ values are derived according to Eqs. 5 and 6.

18